\documentclass[conference]{IEEEtran}
\usepackage{cite}
\usepackage{amsmath,amssymb,amsfonts}
\usepackage{algorithmic}
\usepackage[long]{optidef}
\usepackage{subfigure}
\usepackage{changepage}
\usepackage{graphicx}
\usepackage{textcomp}
\usepackage{xcolor,soul}
\usepackage[linesnumbered,lined,boxed,commentsnumbered,ruled,longend, noend]{algorithm2e}

\SetCommentSty{mycommfont}
\SetKwInput{KwInput}{Input}                
\SetKwInput{KwOutput}{Output}              
\def\BibTeX{{\rm B\kern-.05em{\sc i\kern-.025em b}\kern-.08em T\kern-.1667em\lower.7ex\hbox{E}\kern-.125emX}}

\usepackage{fancyhdr,lipsum}

\fancypagestyle{mahmood}{%
   \fancyhf{} 
   
   \fancyhead[C]{You can use this material personally. Reprinting or republishing this material for the purpose of advertising or promotion, creating new collective works, reselling or redistributing to servers or lists, or using any copyrighted component in other works must adhere to IEEE policy. The DOI will be supplied as soon as it becomes available.}
}%

\makeatletter
\let\ps@IEEEtitlepagestyle\ps@mahmood
\makeatother

\begin{document}

\title{Joint Network Slicing, Routing, and In-Network Computing for Energy-Efficient 6G\\
}

\author{
\centering
\IEEEauthorblockN{Zeinab Sasan*, Masoud Shokrnezhad\textsuperscript{†}, Siavash Khorsandi*, and Tarik Taleb\textsuperscript{†‡}}
\IEEEauthorblockA{* \textit{Amirkabir University of Technology, Tehran, Iran; \{z.sasan, khorsandi\}@aut.ac.ir} \\
\textsuperscript{†} \textit{Oulu University, Oulu, Finland; \{masoud.shokrnezhad, tarik.taleb\}@oulu.fi} 
\\
\textsuperscript{‡} \textit{Ruhr University Bochum, Bochum, Germany; tarik.taleb@rub.de} 
}
}

\maketitle 

\begin{abstract}
To address the evolving landscape of next-generation mobile networks, characterized by an increasing number of connected users, surging traffic demands, and the continuous emergence of new services, a novel communication paradigm is essential. One promising candidate is the integration of network slicing and in-network computing, offering resource isolation, deterministic networking, enhanced resource efficiency, network expansion, and energy conservation. Although prior research has explored resource allocation within network slicing, routing, and in-network computing independently, a comprehensive investigation into their joint approach has been lacking. This paper tackles the joint problem of network slicing, path selection, and the allocation of in-network and cloud computing resources, aiming to maximize the number of accepted users while minimizing energy consumption. First, we introduce a Mixed-Integer Linear Programming (MILP) formulation of the problem and analyze its complexity, proving that the problem is NP-hard. Next, a Water Filling-based Joint Slicing, Routing, and In-Network Computing (WF-JSRIN) heuristic algorithm is proposed to solve it. Finally, a comparative analysis was conducted among WF-JSRIN, a random allocation technique, and two optimal approaches, namely Opt-IN (utilizing in-network computation) and Opt-C (solely relying on cloud node resources). The results emphasize WF-JSRIN's efficiency in delivering highly efficient near-optimal solutions with significantly reduced execution times, solidifying its suitability for practical real-world applications.
\end{abstract}


\section{Introduction}
The forthcoming generation of mobile networks is poised to contend with a substantial surge in traffic volume and user proliferation. Projections indicate that global mobile subscriptions will reach 13.8 billion by 2025 and expand further to 17.1 billion by 2030. Additionally, mobile traffic is anticipated to experience a yearly growth rate of 55\% between 2020 and 2030, as reported by the International Telecommunication Union (ITU) \cite{noauthor_imt_2015}. Furthermore, these next-generation mobile networks are expected to underpin the introduction and development of diverse applications, including mobile holograms, augmented and extended reality, as well as online gaming, among others \cite{yu_toward_2023}. These applications typically necessitate data rates ranging from 1 to 100 Gbps and round-trip delay within the range of 0.1 to 1 ms \cite{kianpisheh_survey_2023}. 
In the context of a network grappling with such burgeoning demand, a substantial user base, and the emergence of novel services, many challenges come to the fore when contemplating the design of next-generation networks.


One fundamental challenge lies in the formulation of a network architecture that can effectively segregate various services and ensure their specific requirements are met. A viable solution to this challenge, particularly in the context of 5G and beyond, is the concept of network slicing, which subdivides a shared infrastructure into multiple distinct logical networks, each tailored to support services with distinct requirements, as elucidated in scholarly works \cite{khan_network_2020, afolabi_network_2018}. Concurrently, another pressing challenge arises from the limited availability of computational and communication resources. Present-day networks predominantly rely on edge-cloud computing for their computational needs, a paradigm that grapples with capacity constraints \cite{kianpisheh_survey_2023}. An emergent solution to this resource scarcity conundrum is in-network computing, which endeavors to address the challenge of computational resource constraints. This strategy involves the utilization of programmable network components such as switches and routers, serving dual roles in facilitating both connectivity and computation \cite{kianpisheh_survey_2023, sapio_-network_2017, benson_-network_2019}.

Network slicing and in-network computing have been extensively investigated in the literature. Sasan ~\textit{et al.}~\cite{sasan_slice-aware_2022} explored resource management in cloud-integrated radio networks to maximize the number of accepted requests. They presented two heuristic algorithms to solve this problem near-optimally. Hu ~\textit{et al.}~\cite{hu_energy-efficient_2021} delved into the realm of in-network computing within next-generation networks, focusing on the processing of micro-service requests on network nodes. Their objective aimed to minimize both link load and energy consumption. To tackle this challenge, the authors employed a multi-objective evolutionary algorithm based on decomposition. Shokrnezhad ~\textit{et al.}~\cite{shokrnezhad_near-optimal_2022, shokrnezhad_double_2023} explored resource allocation within a cloud-network integrated framework. Their objective centered on the minimization of link and processing node costs, and they introduced two heuristic approaches to address this task. Chen ~\textit{et al.}~\cite{chen_network_2020} addressed the problem of network slicing and routing with Virtual Network Function (VNF) placement. Their objective sought to minimize the number of active cloud nodes in the network, thereby reducing energy consumption. The authors streamlined the problem formulation to reduce complexity and facilitate its resolution. Dong ~\textit{et al.}~\cite{dong_intelligent_2022} considered a joint network slicing and routing mechanism with an objective to maximize network utilization and tenant satisfaction. To address resource allocation in this context, they employed the Deep Reinforcement Learning (DRL) method.

While the aforementioned methodologies have demonstrated their effectiveness within specific contexts, their potential suitability for future systems may be limited. Some of these approaches primarily concentrate on resource allocation within network slicing, without incorporating considerations for in-network computing \cite{sasan_slice-aware_2022}. Conversely, there are studies that neglect to capitalize on the computational potential of intermediary nodes through in-network computing technology \cite{shokrnezhad_near-optimal_2022,shokrnezhad_double_2023, chen_network_2020, dong_intelligent_2022}. However, to increase the feasibility of future systems, the most effective strategy involves maximizing the utilization of all available resources through an integrated manner. Furthermore, it is noteworthy that a substantial portion of the existing literature has not adequately addressed the critical issue of energy conservation, while global communication infrastructures have the potential to consume a substantial proportion of the world's electricity, with electricity usage contributing significantly to global greenhouse gas emissions, reaching as high as 23\% by 2030 \cite{noauthor_challenges_nodate}. Consequently, there is an imperative to address the dual challenges of in-network computing and network slicing in a joint manner while concurrently striving to reduce energy consumption.

To fill in the gap in the current literature, our contributions are as follows:
\begin{itemize}
    \item Defining the Mixed-Integer Linear Programming (MILP) formulation of the joint problem of network slicing, routing, and in-network computing with the objective of maximizing the number of accepted users while minimizing energy consumption considering End-to-End (E2E) capacity and quality of service constraints. 
    \item Analyzing the complexity of the optimization problem.
    \item Proposing a Water-Filling-based Joint Slicing, Routing, and In-Network Computing (WF-JSRIN) heuristic algorithm that can yield near-optimal solutions in significantly less time, given the NP-hard nature of the problem.
    \item Assessing the efficiency of WF-JSRIN through a comparative analysis with a random allocation method, as well as two optimal approaches: Opt-IN (which leverages in-network computation) and Opt-C (which relies solely on cloud node resources).
\end{itemize}

The article is structured as follows: In Section \ref{sec_problem_definition}, we present the system model and problem formulation. Section \ref{sec_heuristic_algorithm} introduces the heuristic algorithm designed to provide near-optimal solutions for the problem. Section \ref{sec_performance_evaluation} presents the evaluation of the results obtained from the heuristic algorithm as well as the optimal solution, followed by concluding remarks and potential future directions in Section \ref{sec_conclusion}.

\section{Problem Definition}\label{sec_problem_definition}
In this section, the system model as well as the formulation of the optimization problem are presented

\subsection{System Model}
In this research, we establish a model for the computational and communication network, which is represented as a graph denoted as $G = (\mathbb{V}, \mathbb{E})$. Within this framework, each element $v \in \mathbb{V}$ signifies a network node, while $e \in \mathbb{E}$ denotes network links. Notably, a particular node possesses the highest capacity and is designated as the cloud node, serving as the focal point for routing all network requests. Furthermore, the remaining nodes within the graph $G$ are endowed with programming capabilities, enabling them to actively engage in both computation and communication tasks. Within the confines of graph $G$, each node is distinguished by specific attributes, including computing capacity ($c_v$), delay ($d_v$), and energy consumption cost ($\Psi_v$). Similarly, each link featured in $G$ is associated with parameters delineating capacity ($c_e$), delay ($d_e$), and energy consumption cost ($\gamma_e$). Furthermore, both the cost and energy consumption incurred during the activation of each node are identical and denoted as $\theta_v$. Importantly, it should be emphasized that the energy consumption cost and delay values, whether pertaining to nodes or links, are expressed per unit of data rate.

Moreover, we consider a designated set of network slices, denoted as $\mathbb{M}$. Each of these slices, represented as $m$, encompasses its distinct user ensemble denoted as $\mathbb{U}_m$. Each network slice is strategically isolated from others by receiving an allocated portion of the network's resources. To ensure the fulfillment of the minimum prerequisites for each slice, a fixed resource fraction, represented as $\epsilon_{m}$, is exclusively reserved for each slice, while the remaining resources are distributed equitably among the various slices. Within the confines of each individual slice, individual user requests, denoted as $u_m \in \mathbb{U}_m$, are accompanied by specific criteria, encompassing delay requirements ($d_{u_m}$) and data rate requisites ($r_{u_m}$), which must be meticulously met. It is imperative to underline that, in this study, all conceivable paths for each request originating from the slice's users to the cloud node have been predetermined, with this collection of paths being designated as $\mathbb{P}$. The system model is depicted in Fig. \ref{fig1}.

\begin{figure}
\label{fig1}
\includegraphics[width=3.5in]{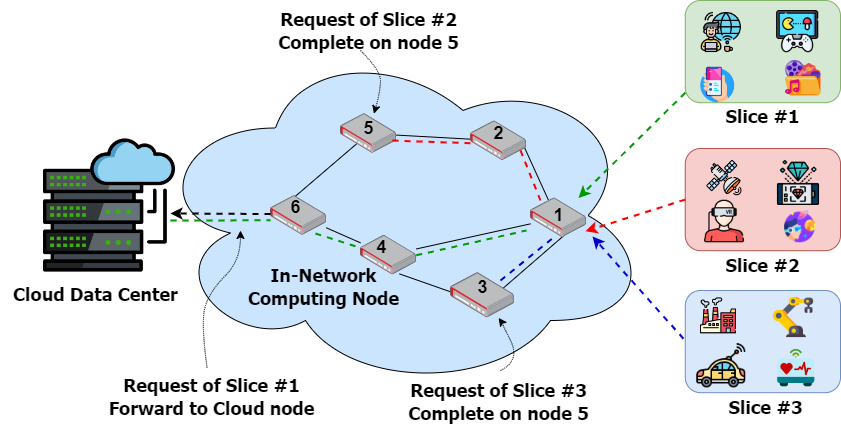}
    \caption{The conceptual overview of joint network slicing, routing, and in-network computing.}
    \vspace{-15pt}
\end{figure}

\subsection{Problem Formulation}
Now, there exists a collection of slices, each containing a distinct set of users, all aiming to gain access to the cloud node in order to fulfill their respective requests. Since traversing all network nodes and executing requests on the cloud node is inefficient in terms of E2E delay and energy consumption, we present the formulation of the network slicing and in-network computation joint problem in this subsection. Our objective here is to allocate the available resources among network slices, with the overarching goal of maximizing the number of supported users while concurrently minimizing energy consumption. The central concept revolves around harnessing the computational capabilities of network nodes to address requests in close proximity to the network edge. This strategy serves the dual purpose of reducing E2E delay for requests and mitigating overall resource utilization, resulting in decreased energy consumption.

The initial two constraints serve to establish a path and the corresponding computational resources for each request. The binary decision variable denoted as $x_{u_m}^p$ is an indicator to determine the selection of path $p$ for user $u_m$ within slice $m$. This binary variable effectively serves as a means of admission control. For the sake of simplicity, it is presumed that each request can opt for only one path. Furthermore, the decision variable $w_{u_m}^{p, v}$ represents the computational capacity that each node $v \in \mathbb{V}_p$ is capable of providing for the request initiated by user $u_m$. Here, $\mathbb{V}_p$ denotes the set of nodes situated along path $p$. In the event that the request initiated by user $u_m$ is routed along path $p$, it becomes imperative to ensure that the total computational capacity allocated to this user's request spans the nodes on path $p$ and aligns with the data rate requirement of the request. This constraint serves to uphold the requirement that the data rate stipulated for $u_m$ remains satisfied.
\begin{align}
& 
\sum_{p \in \mathbb{P}} x_{u_m}^p \leq 1 \quad \forall m, u_m \in \mathbb{M}, \mathbb{U}_m
\tag{C1}
\\
& 
\sum_{v \in \mathbb{V}_p} w_{u_m}^{p, v} = r_{u_m} \cdot x_{u_m}^p \quad \forall m, u_m, p \in \mathbb{M}, \mathbb{U}_m,  \mathbb{P}
\tag{C2}
\end{align}

Having in mind that predetermined paths are established for routing requests from slice users to the cloud node, it is possible that the allocation of computational resources in network nodes may be adequate to satisfy these requests without traversing all links within their designated path. Specifically, C3 serves to identify the specific links that each request occupies along its assigned path. The set $\mathbb{E}_{p,v}$ encompasses all links within path $p$ that lead to node $v$. When path $p$ is selected for routing the request of user $u_m$ and computational processes are conducted on node $v \in \mathbb{V}_p$, it becomes essential to determine the selection of all links $e \in \mathbb{E}_{p,v}$ responsible for directing traffic toward node $v$. To facilitate this determination, we introduce a binary decision variable, denoted as $z_{u_m}^{p, e}$, which serves to indicate whether a given link $e \in \mathbb{E}_{p, v}$ is selected for user $u_m$'s request. In this constraint, the Big M technique is deployed to ensure that $z_{u_m}^{p, e}$ assumes a value of 1 when $w_{u_m}^{p, v}$ is assigned a value, with $\mathcal{M}$ representing a sufficiently large positive constant. Furthermore, if node $v \in \mathbb{V}$ is assigned to at least one user, that node should be turned on. $y_v$ as a binary decision variable is used to indicate whether the node is on or off in C4.
\begin{align}
&
w_{u_m}^{p, v} \leq \mathcal{M} \cdot z_{u_m}^{p, e} \quad
\begin{aligned}
    & \forall m, u_m, p, v, e \in \\
    & \mathbb{M}, \mathbb{U}_m, \mathbb{P}, \mathbb{V}_p, \mathbb{E}_{p, v}
\end{aligned}
\tag{C3}
\\
&
\sum_{m \in \mathbb{M}} \sum_{u_m \in \mathbb{U}_m} \sum_{p \in \mathbb{P}} w_{u_m}^{p, v} \leq \mathcal{M} \cdot y_v \quad \forall v \in \mathbb{V}
\tag{C4}
\end{align}

The concept of slicing in this paper pertains to the allocation of network resources, encompassing both communication resources associated with links and computational resources attributed to network nodes, to each individual slice. This allocation is symbolized as $\lambda_m$ for slice $m$. Constraints C5 and C6 have been devised to rigorously ascertain that the total resources assigned to users within each slice do not surpass the designated allocation $\lambda_m$. These constraints serve the pivotal purpose of enforcing strict isolation between slices, while simultaneously accommodating the finite capacity restrictions imposed on both links and nodes.
\begin{align}
&
\sum_{u_m \in \mathbb{U}_m} \sum_{p \in \mathbb{P}} w_{u_m}^{p, v} \leq \lambda_{m} \cdot c_v \quad \forall m, v \in \mathbb{M}, \mathbb{V} 
\tag{C5}
\\
&
\sum_{u_m \in \mathbb{U}_m} \sum_{p \in \mathbb{P}} z_{u_m}^{p, e} \cdot r_{u_m} \leq \lambda_{m} \cdot c_{e}, \quad \forall m, e \in \mathbb{M}, \mathbb{E}
\tag{C6}
\end{align}
In addition, it is crucial to establish an assurance that the aggregate of allocations among all slices equates to 1. To uphold this requirement, while simultaneously safeguarding each slice's compliance with its minimum prerequisites and preserving its isolation from others to foster equitable treatment, a minimum allocation of $\epsilon_m$ is considered for each slice. Constraints C7 and C8 have been specifically formulated to uphold these constraints.
\begin{align}
&
\lambda_{m} \geq \epsilon_m, \quad  \forall m \in \mathbb{M}
\tag{C7}
\\
&
\sum_{m \in \mathbb{M}} \lambda_{m} =1, \quad 0 \leq \lambda_{m} \leq 1
\tag{C8}
\end{align}

The final constraint, denoted as C9, serves to ensure that the E2E delay encountered by the request initiated by user $u_m$ and allocated to slice $m$ remains equivalent to the summation of its computation and networking delays. Importantly, this cumulative delay must not exceed the stipulated delay requirement ($d_{u_m}$) for the request. The formal expression of this constraint is as follows:
\begin{align}
    \sum_{v \in \mathbb{V}_p} d_v \cdot w_{u_m}^{p, v} + \sum_{e \in \mathbb{E}_p} d_e \cdot z_{u_m}^{p, e}  \cdot r_{u_m} \leq d_{u_m} \quad
    \begin{aligned}
    & \forall u_m, p \\
    & \in \mathbb{U}_m, \mathbb{P}
    \end{aligned}
    \tag{C9}
\end{align}

The final stage in the formulation process involves specifying the objective function (OF). As previously elucidated, the principal aim of this study is to maximize the count of accepted users ($A$) while concurrently minimizing the cost of energy consumption ($B$). To attain a harmonious equilibrium between these objectives, we introduce a coefficient, denoted as $\alpha$, which serves to modulate the relative significance attributed to the number of acceptances in contrast to energy consumption within the objective function. $B$ encompasses several components, namely computations on nodes ($V_{ec}$), node activation ($P_{ec}$), and data transmission over links ($E_{ec}$). In what follows, we provide the definitions of these metrics, A and B, as well as the problem.
\begin{align}\label{CS_PTH1}
    &
    V_{ec} = \sum_{m \in \mathbb{M}} \sum_{u_m \in \mathbb{U}_m} \sum_{p \in \mathbb{P}} \sum_{v \in \mathbb{V}_p} w_{u_m}^{p, v} \cdot \delta_v
    \\
    &
    P_{ec} = \sum_{v \in \mathbb{V}} y_v \cdot \theta_v 
\end{align}
\begin{align}\label{CS_PTH2}
    &
    E_{ec} = \sum_{m \in \mathbb{M}} \sum_{u_m \in \mathbb{U}_m} \sum_{p \in \mathbb{P}} \sum_{e \in \mathbb{E}_p} z_{u_m}^{p, e} \cdot r_{u_m} \cdot \gamma_e
    \\
    &
    A = \sum_{m \in \mathbb{M}} \sum_{u_m \in \mathbb{U}_m} \sum_{p \in \mathbb{P}} x_{u_m}^p
    \\
    &
    B = V_{ec} + E_{ec} + P_{ec}
    \\
    &
    \max_{\lambda, x, y, w, z} \alpha A - B \text{ s.t. } C1-C9 \label{problem}
\end{align}

\subsection{Complexity Analysis}
We establish the NP-hardness of our optimization problem by means of a polynomial-time reduction from the widely recognized Generalized Assignment Problem (GAP), thus firmly substantiating its computational intricacy. Within this reduction, we draw parallels between the agents in GAP, which symbolize resources, and the tasks they undertake, which mirror the requests in our problem. By simplifying the objective function and considering an instance of our problem where all requests can be feasibly accommodated, we elucidate that our problem possesses, at a minimum, the same level of computational complexity as the GAP, further reinforcing its NP-hardness. 

\section{Heuristic Algorithm}\label{sec_heuristic_algorithm}
In this section, we present a comprehensive explanation of the Water-Filling-based Joint Slicing, Routing, and In-Network Computing (WF-JSRIN) algorithm, employed to efficiently address the optimization problem defined in \eqref{problem}. The algorithm is outlined in Algorithm \ref{WF-JSRIN}. Commencing its execution, the algorithm initializes crucial decision variables, denoted as $\lambda$, $x$, $y$, $z$, and $w$, which play a pivotal role in directing the resource allocation process. In order to provide a comprehensive view of resource allocation, the algorithm computes the total data rate ($d_t$), encompassing the data rate requirements of all network slices. For each individual slice, it computes a slice-specific total data rate ($d_m$). Subsequently, the algorithm calculates a slice-specific $\lambda_m$ value, influenced by a predefined parameter $\epsilon_m$, and skillfully adjusted to ensure fair resource allocation among slices, taking into account their respective data rate demands. Within the confines of each slice ($m$), the algorithm initializes node and link capacity variables ($c_{v, m}$ and $c_{e, m}$) tailored to the distinctive characteristics of that slice. As the algorithm proceeds, it embarks on an inner loop for each user ($u_m$). Within this loop, three fundamental variables are initialized: the user's remaining capacity requirement ($\phi_{u_m}$), the cost associated with each path ($\mu_{u_m, p}$), and the delay corresponding to these paths ($\varphi_{u_m, p}$).

\begin{algorithm}[t!]\label{WF-JSRIN}
\caption{WF-JSRIN}
\KwInput{$G$, $\mathbb{M}$, $\epsilon_m \; \forall m \in \mathbb{M}$}
Initialize $\lambda, x, y, z, w$ \\
$d_t$: total data rate of all slices \\

\For{each $m$ in $\mathbb{M}$}
{
    $d_m \gets \text{total data rate of the slice } m$ \\
    $\lambda_m \gets \epsilon_m + (1-\epsilon_m*|\mathbb{M}|)d_m/d_t$
}

\For{each $m$ in $\mathbb{M}$}
{
    $c_{v, m} \gets \text{remaining capacity of nodes for slice } m$ \\
    $c_{e, m} \gets \text{remaining capacity of links for slice } m$ \\
    \For{each $u_m$ in $\mathbb{U}_m$}
    {
        $\phi_{u_m} \gets $ the data rate requirement of $u_m$ \\
        $\mu_{u_m, p} \gets 0$ for each $p$ in $\mathbb{P}$  \\
        $\varphi_{u_m, p} \gets 0$ for each $p$ in $\mathbb{P}$ \\
        \For{each $p$ in user's paths}
        {
            reset $\phi_{u_m}$ \\
            \For{each $v$ in $p$}
            {
                \If{$\phi_{u_m} \neq 0$ \& $c_{v, m} \neq 0$}
                {
                    $y_v \gets 1$ \\
                    $w_{u_m}^{p, v} \gets min\{\phi_{u_m}, c_{v, m}\}$ \\
                    $\Omega = \phi_{u_m} - c_{v, m}$ \\
                    $\phi_{u_m} \gets max\{0, \Omega\}$ \\
                    $c_{v, m} \gets max\{0, -\Omega\}$ \\
                    Update $\varphi_{u_m, p}$ \& $\mu_{u_m, p}$
                }
                \If{$\phi_{u_m} = 0$}
                {
                    \For{each $e$ in $\mathbb{E}_{p, v}$}
                    {
                        \If{$c_{e, m} \geq r_{u_m}$}
                        {
                            $z_{u_m}^{p, e} \gets 1 $ \\
                            $c_{e, m} \gets c_{e, m} - r_{u_m}$ \\
                            Update $\varphi_{u_m, p}$ \& $\mu_{u_m, p}$
                        }
                        \Else
                        {
                            reset allocations and updates
                        }
                    }
                    \If{$\varphi_{u_m, p} > d_{u_m}$}
                    {
                        reset allocations and updates
                    }
                }
            }
            \If{$\phi_{u_m} \neq 0$}
            {
                reset allocations and updates
            }
        }
        $p' \gets argmax_{p \in \mathbb{P}} \; \mu_{u_m, p}$ \\
        $x_{u_m}^{p'} \gets 1$ 
    }
}
return $\lambda, x, y, z, w$
\end{algorithm}

Next, the algorithm systematically iterates through each potential path ($p$) accessible to the user, scrutinizing each node ($v$) along the path. In cases where both the user's remaining capacity and the node's available capacity are non-zero, the algorithm proceeds to allocate the entirety of the available capacity to the user. Following this allocation, it updates the remaining capacity, adjusts the values of $w_{u_m}^{p, v}$ and $y_v$ to reflect this allocation, and increments the cost, encompassing both the energy consumption costs associated with links and nodes, as well as the delay pertaining to the path for the given request. Any remaining portion of the user's requirement is addressed by other nodes along the path in subsequent iterations. If node $v$ represents the specific node where the request's computational capacity requirement is satisfied, the algorithm takes further action to reserve link capacity over the links spanning from the user to node $v$ on path $p$. Correspondingly, it assigns $z_{u_m}^{p, e}$ and updates the cost and the delay parameters. In contrast, if the available resources on the nodes or links along path $p$ fail to adequately support the request, or if the E2E delay surpasses the tolerable delay threshold ($d_{u_m}$), the allocations are disregarded. Finally, the algorithm proceeds to select the path characterized by the minimum cost, designating the chosen path by setting $x_{u_m}^p$ to 1. Notably, the time complexity of this algorithm remains polynomial in relation to the dimensions of the problem, encompassing factors such as the number of slices, users, and network size. 

\section{Performance Evaluation}\label{sec_performance_evaluation}
In this section, we delve into an in-depth exploration of the operational efficiency exhibited by WF-JSRIN, achieving this by conducting a comparative analysis against three distinct benchmarks: Random-JSRIN (R-JSRIN), Opt-IN, and Opt-C. In the context of R-JSRIN, resource allocations are made randomly, devoid of any specific optimization strategy. In stark contrast, Opt-IN and Opt-C adopt a rigorous optimization approach, rendering optimal allocations with the utilization of the SCIP solver, alongside its Python package \cite{maher_pyscipopt_2016}. Notably, while Opt-IN avails itself of the capability for in-network computation, Opt-C is confined to relying solely on the cloud node for resource allocation. The entire computational experimentation is executed on a system equipped with 8 processing cores, 16 GB of memory, and a 64-bit operating system. The intricate details of the algorithm's implementation parameters are meticulously documented in Table \ref{tabel1}.

The execution times of the employed methods can be found in Table \ref{tabel2}. Notably, the execution times for both WF-JSRIN and R-JSRIN consistently exhibit durations of less than one second, remaining remarkably stable even as the number of users experiences growth. In contrast, when contemplating the Opt-C method, which exclusively relies upon the cloud node to manage computation requests, devoid of any distribution to alternative nodes, the execution time registers a noticeable reduction when compared to Opt-IN. Conversely, the execution time associated with Opt-IN exhibits an exponential increase as the user count escalates. This escalation is attributed to the intricate decision-making process entailed in resource distribution for various users spanning diverse nodes and paths. These empirical findings distinctly underscore that within medium and large-scale networks, the adoption of optimal methodologies entails considerable time overhead. In such scenarios, the preference should be directed towards the employment of heuristic algorithms, which prove to be more time-efficient.

In Fig. \ref{fig1}-A and -B, we examine the cumulative cost, which includes energy consumption costs for both nodes and links, and bandwidth usage as we increment the number of users in each slice. It's evident that regardless of the number of users, Opt-C consistently incurs higher costs and greater bandwidth usage compared to other algorithms. This discrepancy arises because requests in the Opt-C approach traverse more links to reach the cloud, resulting in increased energy consumption and resource usage compared to in-network computing. Additionally, processing at the cloud node is costlier than at intermediary nodes within the network due to the cloud node's superior processing capacity and speed. R-JSRIN also consumes more bandwidth and incurs higher costs than Opt-IN and WF-JSRIN due to the possibility of selecting more distant nodes for request distribution. Conversely, WF-JSRIN utilizes less bandwidth than Opt-IN by initiating computation distribution from closer nodes (resulting in fewer links being traversed) for small numbers of users. As the number of users increases, it navigates more links, causing its bandwidth usage to surpass that of Opt-IN. However, it achieves very similar results to Opt-IN, which aims to minimize total energy consumption and maximize the number of accepted users, demonstrating its near-optimal efficiency even for large numbers of users.

\begin{table}[!t]
\centering
\caption{Simulation parameters}
\vspace{-5pt}
\label{tabel1}
\begin{tabular}{ |c|c| } 
\hline
Parameters & Values  \\
\hline
number of nodes in graph $G$ & 12 \\
cloud node capacity & $\sim U[200, 300]$  \\ 
nodes capacity ($c_v$) & $\sim U[15, 50]$  \\ 
links capacity ($c_e$) & $\sim U[100, 200]$ \\ 
links \& nodes delay ($d_e$, $d_v$) & $\sim U[1, 3]$  \\
links \& nodes cost ($\theta_e$, $\delta_v$, and $\theta_v$) & $\sim U[1, 5]$  \\
data rate of users ($r_{u_m}$) & $\sim U[10, 20]$ \\
delay of users ($d_{u_m}$) & $\sim U[500, 1000]$ \\
\hline
\end{tabular}
\vspace{-5pt}
\end{table}

\begin{table}[!t]
    \centering
    \label{tabel2}
    \caption{Solving time vs. the number of users (second)}
    \vspace{-5pt}
    \begin{tabular}{|c|c|c|c|c|}
    \hline
        \textbf{Users Num} & \textbf{Opt-IN} & \textbf{Opt-C} & \textbf{WF-JSRIN} & \textbf{R-JSRIN} \\ \hline \hline
        3 & 0.528 & 0.0305 & 0.015 & 0.015 \\ \hline 
        6 & 0.530 & 0.060 & 0.018 & 0.030 \\ \hline
        9 & 4.015 & 0.120 & 0.034 & 0.038 \\ \hline
        12 & 101.874 & 0.176 & 0.054 & 0.062 \\ \hline
        15 & 129.460 & 0.263 & 0.089 & 0.087 \\ \hline
        18 & 405.497 & 0.383 & 0.124 & 0.132 \\ \hline
        21 & 3474.588 & 0.574 & 0.183 & 0.180 \\ \hline
        24 & 6543.679 & 2.287 & 0.234 & 0.217 \\ \hline
        27 & 9612.770 & 2.856 & 0.245 & 0.228 \\ \hline
        30 & 12681.8615 & 3.337 & 0.361 & 0.352 \\ \hline
    \end{tabular}
    \vspace{-10pt}
    \label{tab2}
\end{table}

\begin{figure*}[t!]\centering
\includegraphics[width=6in]{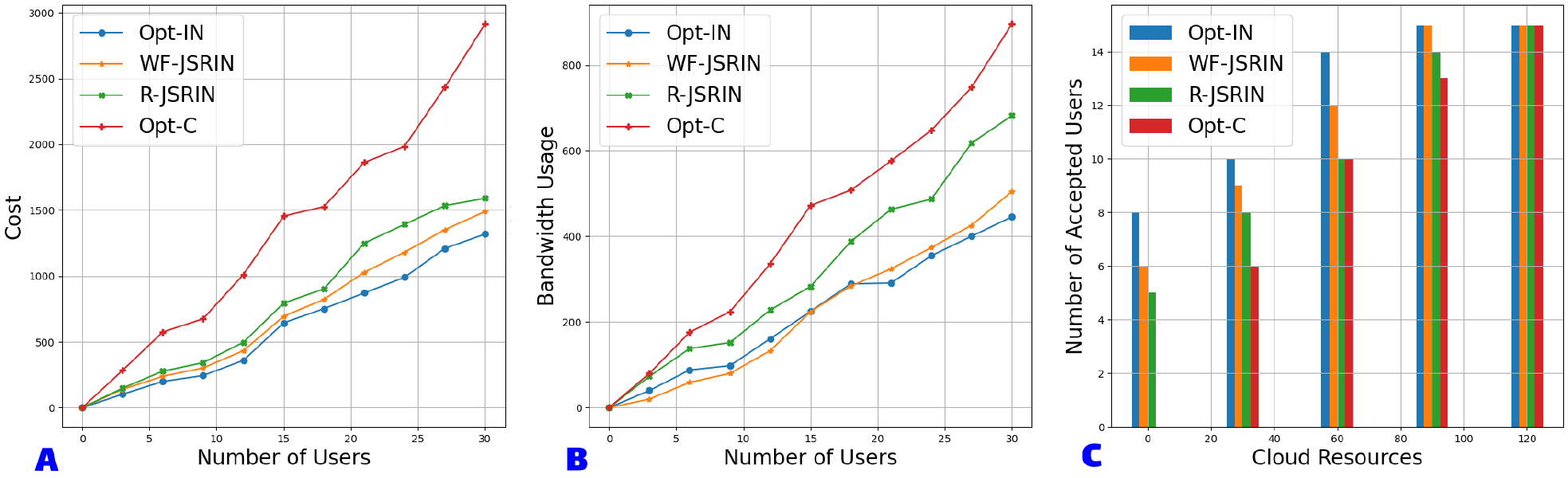}
\vspace{-10pt}
  \caption{The total cost and bandwidth usage vs. the number of users (A \& B), and the number of accepted users vs. the link capacity (C).}
  \vspace{-12pt}
  \label{fig1}
\end{figure*}

In Fig. \ref{fig1}-C, we examine the impact of changing the cloud node's capacity on the number of accepted users, leveraging three slices, each accommodating five users. Notably, augmenting cloud node resources leads to a commensurate increase in users accepted by the Opt-C approach. Remarkably, even in scenarios with limited cloud resources, all approaches leveraging in-network computation outperform Opt-C, with WF-JSRIN closely resembling Opt-IN's performance. We also scrutinize the impact of varying link capacities on user acceptance across different algorithms, yielding outcomes akin to those in Fig. \ref{fig1}-C. In situations characterized by constrained link capacity, in-network computation excels at accommodating more users by circumventing the transmission of a multitude of requests to the cloud node, thus optimizing resource utilization. Conversely, with expanded link capacity, Opt-IN and Opt-C achieve an equitable distribution of accepted users. However, when contemplating the R-JSRIN algorithm, the node selected for request computation may be positioned distantly, necessitating traversal through numerous links. In cases of inadequate link capacity rendering the path unusable for R-JSRIN, its performance lags behind that of WF-JSRIN. These scenarios collectively underscore the exceptional efficiency demonstrated by WF-JSRIN when navigating diverse resource capacity scenarios.

\section{Conclusion}\label{sec_conclusion}
This paper investigated resource allocation within the context of next-generation mobile networks, encompassing the complex task of joint decision-making involving network slicing, routing, and in-network computing. Given the NP-hard nature of this problem, we introduced WF-JSRIN, a Water Filling-based heuristic algorithm, and examined its efficiency against random and optimal solutions, including Opt-IN (utilizing in-network computation) and Opt-C (solely relying on cloud node resources). Our results highlight WF-JSRIN's ability to provide highly efficient near-optimal solutions, demonstrating superior performance in terms of cost, bandwidth utilization, and user acceptance. Furthermore, WF-JSRIN distinguishes itself with remarkably shorter execution times, rendering it a more pragmatic choice for real-world applications. 

Our ongoing research aims to integrate our prior work \cite{farhoudi2023qos} with this study to address the dynamic nature of the problem, considering challenges posed by user mobility and evolving availability using machine learning techniques. Exploring the incorporation of radio access resource allocation, as discussed in our previous work \cite{shokrnezhad_scalable_2023}, and addressing integrated Access and Backhaul (IAB)-enabled E2E resource allocation represent another avenue for future research. Additionally, the integration of slicing for Quantum Networks (QNs) \cite{prados-garzon_deterministic_2023} with in-network computation offers opportunities for future applications, such as the metaverse.

\vspace{-5pt}
\section*{Acknowledgment}
This research work was also conducted at ICTFICIAL Oy. It is partially supported by the European Union’s Horizon 2020 Research and Innovation Program through the aerOS project under Grant No. 101069732; the Business Finland 6Bridge 6Core project under Grant No. 8410/31/2022; the European Union’s HE research and innovation programe HORIZON-JUSNS-2022 under the 6GSandbox project (Grant No. 101096328); and and the Research Council of Finland 6G Flagship Programme under Grant No. 346208. The paper reflects only the authors’ views. The Commission is not responsible for any use that may be made of the information it contains.
\vspace{-5pt}


\bibliographystyle{IEEEtran}
\bibliography{IEEEabrv,cas-refs}

\begin{thebibliography}{10}
\providecommand{\url}[1]{#1}
\csname url@samestyle\endcsname
\providecommand{\newblock}{\relax}
\providecommand{\bibinfo}[2]{#2}
\providecommand{\BIBentrySTDinterwordspacing}{\spaceskip=0pt\relax}
\providecommand{\BIBentryALTinterwordstretchfactor}{4}
\providecommand{\BIBentryALTinterwordspacing}{\spaceskip=\fontdimen2\font plus
\BIBentryALTinterwordstretchfactor\fontdimen3\font minus \fontdimen4\font\relax}
\providecommand{\BIBforeignlanguage}[2]{{%
\expandafter\ifx\csname l@#1\endcsname\relax
\typeout{** WARNING: IEEEtran.bst: No hyphenation pattern has been}%
\typeout{** loaded for the language `#1'. Using the pattern for}%
\typeout{** the default language instead.}%
\else
\language=\csname l@#1\endcsname
\fi
#2}}
\providecommand{\BIBdecl}{\relax}
\BIBdecl

\bibitem{noauthor_imt_2015}
\BIBentryALTinterwordspacing
``{IMT} {Traffic} {Estimates} for the {Years} 2020 to 2030,'' International Telecommunication Union (ITU), Tech. Rep. ITU-R M.2370-0, Jul. 2015. [Online]. Available: \url{https://www.itu.int/pub/R-REP-M.2370-2015}
\BIBentrySTDinterwordspacing

\bibitem{yu_toward_2023}
H.~Yu, M.~Shokrnezhad \emph{et~al.}, ``Toward {6G}-{Based} {Metaverse}: {Supporting} {Highly}-{Dynamic} {Deterministic} {Multi}-{User} {Extended} {Reality} {Services},'' \emph{IEEE Network}, vol.~37, no.~4, pp. 30--38, Jul. 2023.

\bibitem{kianpisheh_survey_2023}
S.~Kianpisheh and T.~Taleb, ``A {Survey} on {In}-{Network} {Computing}: {Programmable} {Data} {Plane} and {Technology} {Specific} {Applications},'' \emph{IEEE Communications Surveys \& Tutorials}, vol.~25, no.~1, pp. 701--761, 2023.

\bibitem{khan_network_2020}
L.~U. Khan, I.~Yaqoob \emph{et~al.}, ``Network {Slicing}: {Recent} {Advances}, {Taxonomy}, {Requirements}, and {Open} {Research} {Challenges},'' \emph{IEEE Access}, vol.~8, pp. 36\,009--36\,028, 2020.

\bibitem{afolabi_network_2018}
I.~Afolabi, T.~Taleb \emph{et~al.}, ``Network {Slicing} and {Softwarization}: {A} {Survey} on {Principles}, {Enabling} {Technologies}, and {Solutions},'' \emph{IEEE Communications Surveys \& Tutorials}, vol.~20, no.~3, pp. 2429--2453, 2018.

\bibitem{sapio_-network_2017}
A.~Sapio, I.~Abdelaziz \emph{et~al.}, ``In-{Network} {Computation} is a {Dumb} {Idea} {Whose} {Time} {Has} {Come},'' in \emph{Proceedings of the 16th {ACM} {Workshop} on {Hot} {Topics} in {Networks}}, ser. {HotNets}-{XVI}.\hskip 1em plus 0.5em minus 0.4em\relax New York, NY, USA: Association for Computing Machinery, Nov. 2017, pp. 150--156.

\bibitem{benson_-network_2019}
T.~A. Benson, ``In-{Network} {Compute}: {Considered} {Armed} and {Dangerous},'' in \emph{Proceedings of the {Workshop} on {Hot} {Topics} in {Operating} {Systems}}, ser. {HotOS} '19.\hskip 1em plus 0.5em minus 0.4em\relax New York, NY, USA: Association for Computing Machinery, May 2019, pp. 216--224.

\bibitem{sasan_slice-aware_2022}
Z.~Sasan and S.~Khorsandi, ``Slice-{Aware} {Resource} {Calendaring} in {Cloud}-based {Radio} {Access} {Networks},'' in \emph{2022 30th {International} {Conference} on {Electrical} {Engineering} ({ICEE})}, May 2022, pp. 1005--1009.

\bibitem{hu_energy-efficient_2021}
N.~Hu, Z.~Tian, X.~Du, and M.~Guizani, ``An {Energy}-{Efficient} {In}-{Network} {Computing} {Paradigm} for {6G},'' \emph{IEEE Transactions on Green Communications and Networking}, vol.~5, no.~4, pp. 1722--1733, Dec. 2021.

\bibitem{shokrnezhad_near-optimal_2022}
M.~Shokrnezhad and T.~Taleb, ``Near-optimal {Cloud}-{Network} {Integrated} {Resource} {Allocation} for {Latency}-{Sensitive} {B5G},'' in \emph{{GLOBECOM} 2022 - 2022 {IEEE} {Global} {Communications} {Conference}}, Dec. 2022, pp. 4498--4503.

\bibitem{shokrnezhad_double_2023}
M.~Shokrnezhad, T.~Taleb \emph{et~al.}, ``Double {Deep} {Q}-{Learning}-based {Path} {Selection} and {Service} {Placement} for {Latency}-{Sensitive} {Beyond} {5G} {Applications},'' \emph{IEEE Transactions on Mobile Computing}, pp. 1--14, 2023.

\bibitem{chen_network_2020}
W.-K. Chen, Y.-F. Liu \emph{et~al.}, ``Network {Slicing} for {Service}-{Oriented} {Networks} with {Flexible} {Routing} and {Guaranteed} {E2E} {Latency},'' in \emph{2020 {IEEE} 21st {International} {Workshop} on {Signal} {Processing} {Advances} in {Wireless} {Communications} ({SPAWC})}, May 2020, pp. 1--5.

\bibitem{dong_intelligent_2022}
T.~Dong, Z.~Zhuang \emph{et~al.}, ``Intelligent {Joint} {Network} {Slicing} and {Routing} via {GCN}-{Powered} {Multi}-{Task} {Deep} {Reinforcement} {Learning},'' \emph{IEEE Transactions on Cognitive Communications and Networking}, vol.~8, no.~2, pp. 1269--1286, Jun. 2022.

\bibitem{noauthor_challenges_nodate}
\BIBentryALTinterwordspacing
``Challenges {on} {Global} {Electricity} {Usage} of {Communication} {Technology}: {Trends} to 2030.'' [Online]. Available: \url{https://www.mdpi.com/2078-1547/6/1/117}
\BIBentrySTDinterwordspacing

\bibitem{maher_pyscipopt_2016}
S.~Maher, M.~Miltenberger \emph{et~al.}, ``\BIBforeignlanguage{en}{{PySCIPOpt}: {Mathematical} {Programming} in {Python} with the {SCIP} {Optimization} {Suite}},'' in \emph{\BIBforeignlanguage{en}{Mathematical {Software} – {ICMS} 2016}}, ser. Lecture {Notes} in {Computer} {Science}.\hskip 1em plus 0.5em minus 0.4em\relax Cham: Springer International Publishing, 2016, pp. 301--307.

\bibitem{farhoudi2023qos}
M.~Farhoudi, M.~Shokrnezhad \emph{et~al.}, ``Qos-aware service prediction and orchestration in cloud-network integrated beyond 5g,'' \emph{arXiv preprint arXiv:2309.10185}, 2023.

\bibitem{shokrnezhad_scalable_2023}
M.~Shokrnezhad, S.~Khorsandi \emph{et~al.}, ``A {Scalable} {Communication} {Model} to {Realize} {Integrated} {Access} and {Backhaul} ({IAB}) in {5G},'' in \emph{{ICC} 2023 - {IEEE} {International} {Conference} on {Communications}}, May 2023, pp. 1350--1356.

\bibitem{prados-garzon_deterministic_2023}
J.~Prados-Garzon, T.~Taleb \emph{et~al.}, ``Deterministic {6GB}-{Assisted} {Quantum} {Networks} with {Slicing} {Support}: {A} {New} {6GB} {Use} {Case},'' \emph{IEEE Network}, pp. 1--1, 2023.

\end{thebibliography}

\end{document}